\documentclass[twocolumn,showpacs,floats,superscriptaddress,aps,prl]{revtex4-1}
\usepackage{amsfonts}
\usepackage{amssymb}
\usepackage{amsmath}
\usepackage{graphicx}
\usepackage{bm}
\usepackage{color}
\usepackage{float}

\usepackage{calc}
\usepackage{ifthen}
\usepackage{mwe}
\usepackage{subfigure}

\usepackage{epstopdf}
\usepackage{epsfig}
\usepackage{braket}
\usepackage{hyperref} 
\hypersetup{
    colorlinks = true,
    linkcolor = {blue},
    citecolor = {blue},
}

\begin{document}
\raggedbottom


\title{Detuning-modulated composite pulses for high-fidelity robust quantum control}

\date{\today}

\author{Elica Kyoseva}
\email{elkyoseva@gmail.com}
\thanks{These authors contributed equally}
\affiliation{Institute of Solid State Physics, Bulgarian Academy of Sciences, 72 Tsarigradsko Chaussee, 1784 Sofia, Bulgaria}
\affiliation{Condensed Matter Physics Department, School of Physics and Astronomy, Tel Aviv University, Tel Aviv 69978, Israel}
\affiliation{Center for Light-Matter Interaction, Tel Aviv University, Tel Aviv 69978, Israel}

\author{Hadar Greener}
\thanks{These authors contributed equally}
\affiliation{Condensed Matter Physics Department, School of Physics and Astronomy, Tel Aviv University, Tel Aviv 69978, Israel}
\affiliation{Center for Light-Matter Interaction, Tel Aviv University, Tel Aviv 69978, Israel}

\author{Haim Suchowski}%
\affiliation{Condensed Matter Physics Department, School of Physics and Astronomy, Tel Aviv University, Tel Aviv 69978, Israel}
\affiliation{Center for Light-Matter Interaction, Tel Aviv University, Tel Aviv 69978, Israel}

\begin{abstract}
We present a novel set of composite pulses, which use the off-resonant detuning as their control parameter. 
\end{abstract}

                             

\begin{abstract}
We introduce a novel control method for robust quantum information processing suited for quantum integrated photonics. We utilize off-resonant detunings as control parameters to derive a new family of composite pulses for high-fidelity population transfer within the quantum error threshold. By design, our detuning-modulated $N$-piece composite sequences correct for any control inaccuracies including pulse strength, duration, resonance offsets errors, Stark shifts, unwanted frequency chirp, etc. We reveal the symmetries of the pulses that allow for straightforward scaling with minimal pulse overhead for an arbitrary $N$. Furthermore, we implement the composite solutions in coupled waveguides allowing a complete light transfer that is robust to fabrication errors.
\end{abstract}

\maketitle


Quantum information processing (QIP) relies on high-fidelity quantum state preparation and transfer. This presents a challenge in practical realizations of QIP where the admissible error of quantum operations is smaller than $10^{-4}$ \cite{QuantumInfo1}. Thus, even small systematic errors, i.e., due to imperfections in fabrication or in the experimental control knobs, reduce the fidelity of state transfer below the fault-tolerant threshold. A powerful tool to correct for systematic errors are composite pulses (CPs), which were initially developed in the field of nuclear magnetic resonance \cite{CP1,CarrSpinEcho,CP2,CP3,CP4,CP5,CP6,CP7}. A composite pulse is a sequence of pulses with different areas and/or phases that implement accurate and robust quantum gates. To this end, CPs are designed for resonant or adiabatic interactions with complex coupling parameters  \cite{VitanovSmooth,Torosov:adiabatic,GenovPRL} and were successfully used to achieve complete population transfer (CPT) in quantum systems in both radiofrequency (rf) and ultrashort pulsed excitations \cite{KeelerNMR}. 

More recently, CPs were applied in many physical realizations of QIP including trapped ions \cite{PhysRevA.93.032340} and atomic systems \cite{Elica:passband,VitanovUltraBB}, and also to achieve accuracy in matching higher harmonic generation processes \cite{Rangelov:14} and in designing polarization rotators \cite{CPPolarizer,Dimova16}. Another promising candidate for advancing QIP technologies is integrated photonic circuits due to their scalability and on-chip integration capacity \cite{Politi1221, MultiphotonEntanWG, PSTPeruzzo}. However, the fidelity of operations remains below the QIP threshold due to unavoidable fabrication errors. CPs have not been previously used to correct for such errors as existing sequences require control of the phase of the coupling, which in integrated photonic circuits is a real parameter. The present research is the first to address this limitation and to derive CPs that can be used in any qubit architecture including integrated photonic systems.

\begin{figure}
\includegraphics[scale=0.25]{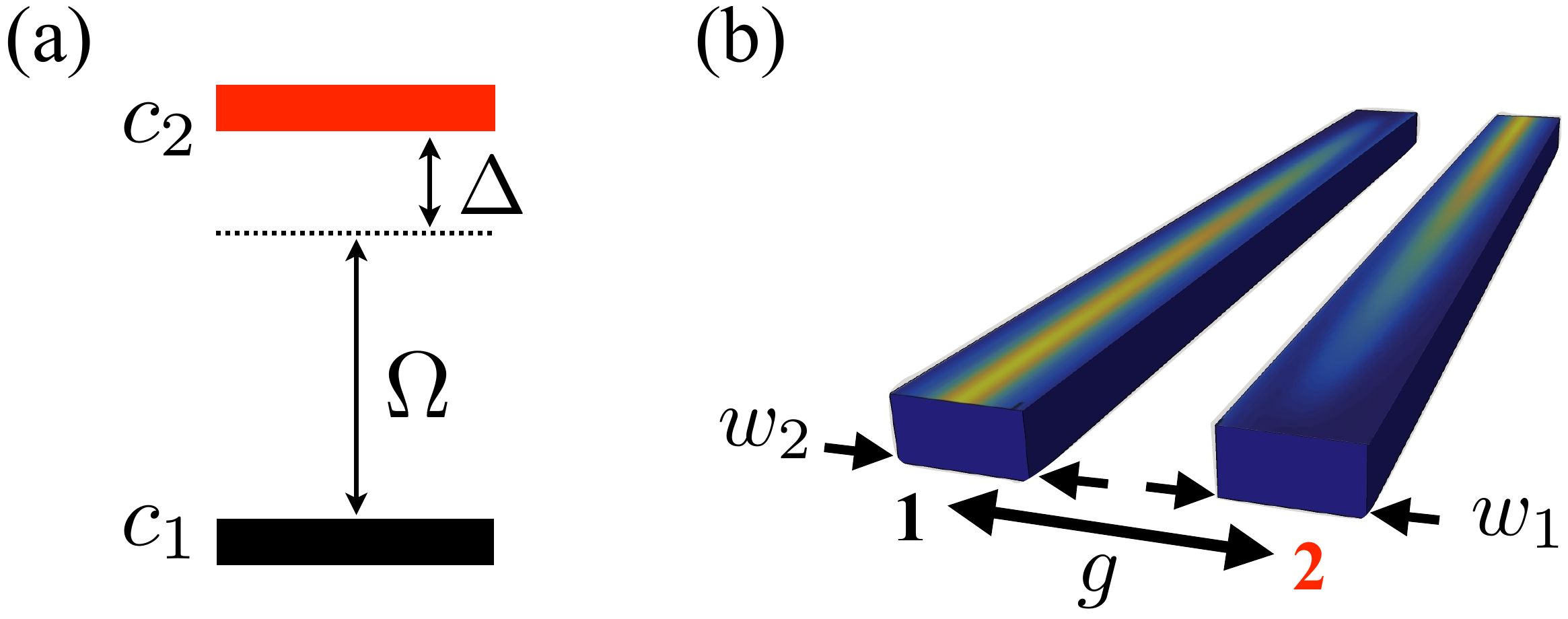}
\caption{(color online) Qubit dynamics. (a) A two-state atomic system with ground
state $c_{1}$ and excited state $c_{2}$ coupled via Rabi frequency $\Omega$ and detuning $\Delta$. (b) An optical directional coupler comprising two waveguides of widths $w_{i}$ at a distance $g$ can be described as a two-state quantum system.}
\label{fig1}
\end{figure}

In this Letter, we introduce the first composite sequences designed for off-resonant complete and robust qubit inversion without any constraints on the coupling strengths. We achieve the desired high-fidelity population transfer by suitably choosing the detuning parameters while maintaining constant coupling. The presented general approach to derive detuning-modulated composite pulses of an arbitrary length $N$ has a minimal pulse overhead and robust transfer is realized even for $N=2$. In our analysis we consider a generic coupled two-state quantum system which has many possible physical realizations including atomic and photonic (shown in Fig. \ref{fig1}). We show that our solution is inherently stable to inaccuracies in all systematic parameters --- coupling strength, coupling duration and resonance offsets --- and achieves fidelities well above the QIP gate error threshold making it a cornerstone for high-fidelity quantum operations for QIP.  Finally, we layout the general recipe to implement the presented detuning-modulated composite sequences in integrated photonic systems for broadband high fidelity optical switching. 

{\it Detuning-modulated composite pulses.}
The time evolution of a qubit system \{$|1\rangle, |2\rangle$\}, shown in Fig. \ref{fig1} (a), driven coherently by an external electromagnetic field is governed by the Schr\"{o}dinger equation
\begin{equation}
i \hbar \partial_t \left[ \begin{array}{c}
c_1(t) \\
c_2(t) \end{array} \right] = \frac{\hbar}{2} 
\left[ \begin{array}{cc}
-\Delta(t) & \Omega(t) \\
\Omega^*(t) & \Delta(t) \end{array} \right] 
\left[ \begin{array}{c}
c_1(t) \\
c_2(t) \end{array} \right] .
\label{Sch}
\end{equation}
Here, $[c_1(t),c_2(t)]^{\textnormal{T}}$ is the probability amplitudes vector, $\Omega(t)$ is the Rabi frequency of the transition, and $\Delta(t) = \omega_0 - \omega$ is the real-valued detuning between the laser frequency $\omega$ and the Bohr transition frequency of the qubit $\omega_0$. In what follows, 
we assume $\Omega(t)$ and $\Delta(t)$ real and constant, which is well-suited for the foreseen implementation of our results in coupled waveguides and in optical elements for generating higher harmonics. 

The unitary propagator of the time evolution governed by Eq. \eqref{Sch} is found according to $U(t,0) = e^{-i/\hbar \int_{0}^{t} H(t) dt}$, 
\begin{equation}
U(\delta t) = \left[ \begin{array}{cc}
\cos \left(\frac{A}{2} \right)+i\frac{ \Delta}{\Omega_g} \sin \left(\frac{A}{2} \right) & -i\frac{ \Omega}{ \Omega_g } \sin \left(\frac{A}{2} \right) \\
- i\frac{\Omega}{ \Omega_g } \sin \left( \frac{A}{2} \right) & \cos \left(\frac{A}{2} \right) - i\frac{ \Delta}{\Omega_g} \sin \left( \frac{A}{2} \right) \end{array} \right].
\label{U}
\end{equation}
Here, $\Omega_g = \sqrt{ \Omega^2 + \Delta^2}$ is the generalized Rabi frequency and $A =  \Omega_g \delta t$ is the pulse area with $\delta t =(t-t_0) $ being the pulse duration. The propagator $U(\delta t)$ evolves the state of the qubit from the initial time $t_0$ to the final time $t$ according to $\mathbf{c}(t) = U(\delta t) \mathbf{c}(t_0)$. If the initial state of the qubit at $t_0$ is $|1\rangle$, the population of the excited state $|2\rangle$ at time $t$ is found by the modulus squared of the off-diagonal propagator element $|U_{12}(\delta t )|^2$. 

We assume a composite pulse sequence comprising $N$ individual off-resonant pulses with Rabi frequencies $\Omega_n$ and detunings $\Delta_n$. Given the individual pulse propagator $U_n(\delta t_n)$ from Eq. \eqref{U}, the propagator for the total composite pulse sequence is given by the product 
\begin{equation}
U^{(N)}(T,0) = U_{N}(\delta t_N)\;U_{N-1}(\delta t_{N-1}) \dots U_{1}(\delta t_1),
\label{U_tot}
\end{equation}
where $\delta t_n = (t_{n} - t_{n-1})$ is the duration of the $n^{\textnormal{th}}$ pulse ($t_0 = 0$ and $t_N \equiv T$). 
Without loss of generality, we focus on the case of individual $\pi$-pulses, i.e., $A_n = A = \pi$, which is easily realized by setting the pulse durations according to $\delta t_n = \pi/\Omega_{g,n} $.

{\it $N$-pulse broadband composite sequences.} We require that the composite sequence produces complete qubit flip at the end of the evolution, that is, the modulus squared of the off-diagonal element from Eq. \eqref{U_tot} is $|U_{12}^{(N)}(T,0) |^2=1$. See Supplemental Material at [.....] for the exact form of the propagator element for an arbitrary $N$. We use the set of detuning values $ \{ \Delta_n \} $ as control parameters and find that for a qubit flip they need to fulfill a general analytical condition depending on the value of $N$. For an even number of pulses $N=2n$ that is
\begin{flalign} 
& 1  + \sum_{i< j =1}^{2n} (-1)^{i+j+1} \frac{\Delta_i}{\Omega_i}  \frac{\Delta_j}{\Omega_j} + \dotso + & \nonumber \\
& +  \sum_{i < \dotso < m = 1}^{2n} (-1)^{i+\dotso +m +1} \underbrace{ \frac{\Delta_i}{\Omega_i} \cdots  \frac{\Delta_m}{\Omega_m}}_{2n\, \textnormal{times}}  = 0, &
\label{cpt2}
\end{flalign}
while for an odd number of pulses $N=(2n+1)$ it is,
\begin{flalign} 
& \sum_{i=1}^{2n+1}(-1)^{i+1} \frac{\Delta_i}{\Omega_i} +  \sum_{i < j < k =1}^{2n+1} (-1)^{i+j+k} \frac{\Delta_i}{\Omega_i} \frac{\Delta_j}{\Omega_j} \frac{\Delta_k}{\Omega_k}  &  \nonumber \\ 
&+ \dotso + \sum_{i < \dotso < m = 1}^{2n+1} (-1)^{i+\dotso +m +1} \underbrace{\frac{\Delta_i}{\Omega_i} \cdots \frac{\Delta_m}{\Omega_m}}_{(2n+1) \, \textnormal{times}} =0. &
\label{cpt3}
\end{flalign}

\begin{figure}[tbp]
\includegraphics[scale=0.4]{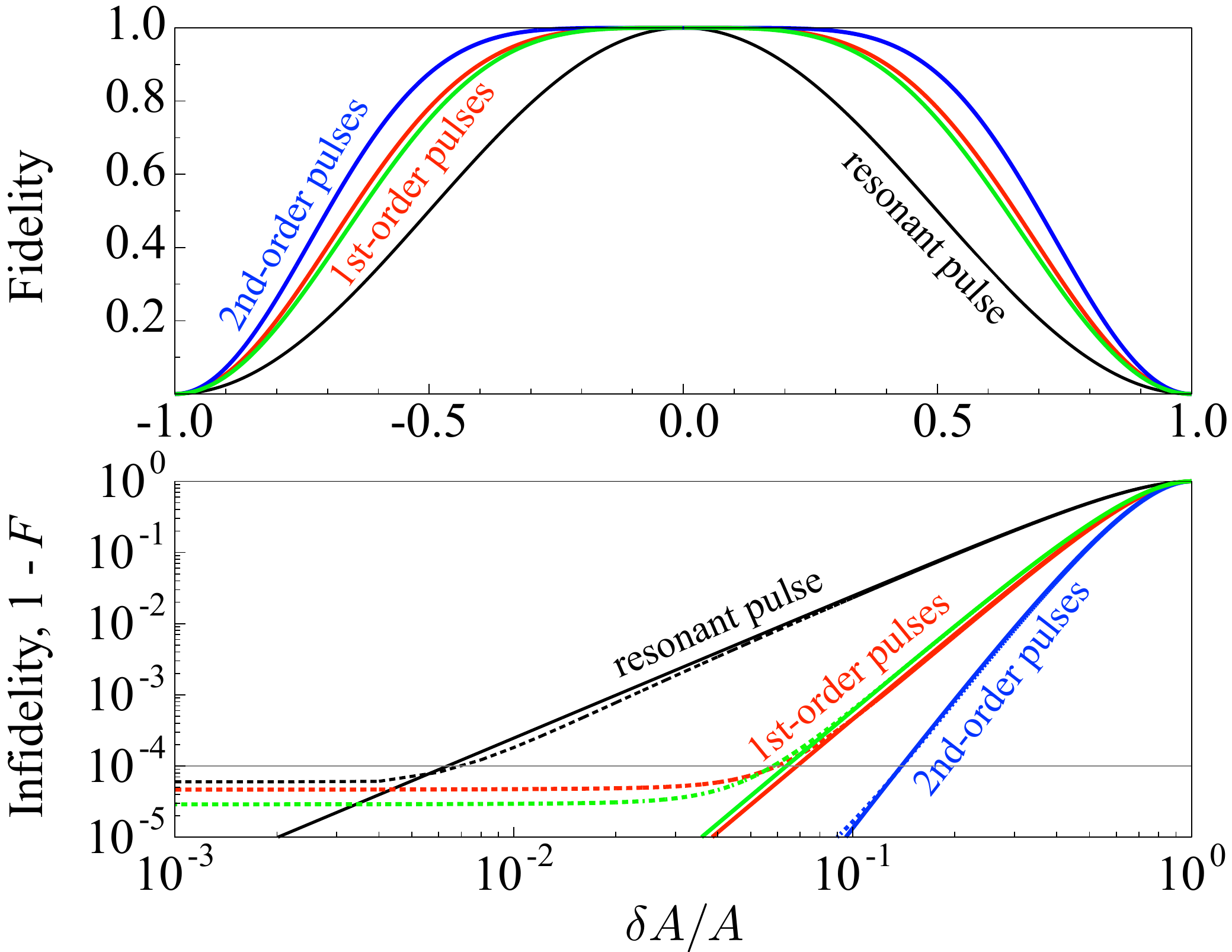}
\caption[justification=justified]{(color online) Fidelity of the detuning-modulated composite sequences vs errors in the pulse area $\delta A/A$. (top) Fidelity of the first-order CPs with $N=2,3$ (green, red) constituent pulses, and of the second-order sequence with $N=3$ (blue). (bottom) Infidelity of the CPs from (top) with (dashed curves) and without (solid curves) Gaussian errors of 10\% in $A$ averaged over 100 times. The fidelity of a resonant pulse is shown in black.}
\label{fig2}
\end{figure}

A robust composite sequence that corrects for imperfections in the pulse area exhibits the signature flat-top profile at $A=\pi$ as shown in Fig. \ref{fig2} (top). We achieve this by taking the partial derivatives of the off-diagonal element from Eq. \eqref{U_tot}, $\frac{\partial^k}{\partial A^k} |U_{12}^{(N)}(T,0)|^2$, at $A=\pi$ and consecutively nullifying them ($k=1,2,\dots$) \cite{VitanovSmooth}. The odd derivatives are always equal to zero and thus for a {\it first-order CP} we need to nullify the second derivative, while for a {\it second-order CP} -- the second and the fourth derivatives simultaneously. Note, that in contrast to previous works \cite{VitanovSmooth,Torosov:adiabatic}, the pulse area $A$ for an off-resonant pulse is a function of all systematic parameters---pulse duration, amplitude, and detuning---and thus, the detuning-modulated composite pulses presented here are {\it robust against all systematic errors by design}. 

In the following we provide analytical solutions for broadband pulse sequences of arbitrary lengths $N$ with equal couplings ($\Omega_n = \Omega$) by utilizing symmetry rules for the detuning values of the individual pulses. The presented sequences will be straightforward to realize in NMR and in coupled waveguides qubits. We note that other sets of solutions that do not fulfill the given symmetry rules exist and can be obtained by numerically nullifying the partial derivatives. However, this task proves computationally challenging even for short composite sequences of length $N=3$ due to the complicated form of the propagator element and its derivatives. Thus, the uncovered pulse symmetries represent a powerful analytical quantum control tool and allow finding sequences of arbitrary lengths in a straightforward manner. 

{\it First-order composite pulses: Sign-alternating $\Delta $.} We consider individual detuning parameters that are equal in magnitude and alternating in sign, i.e., $\Delta_i = - \Delta_{i+1} \equiv \Delta$ for $i=(1,\dots ,N-1)$. Then, the CPT conditions Eqs. \eqref{cpt2} and \eqref{cpt3} can be combined and rewritten as the polynomial 
\begin{eqnarray}
\sum_{s=0}^{n} (-1)^s \Omega^{2s} \binom{N}{N - 2s} \Delta^{N-2s}  = 0,
\label{poly}
\end{eqnarray}
which is valid for both even $N=2n$ and odd $N = (2n+1)$ sequences. The roots of this polynomial provide the values of $\Delta$ for which a complete bit flip is achieved and moreover $\frac{\partial^2}{\partial A^2} |U_{12}^{(N)}|^2 $ at $A=\pi$ is nullified. For a flat-top broadband composite sequence we choose the root that minimizes the fourth derivative $\frac{\partial^4}{\partial A^4} |U_{12}^{(N)}|^2 $ at $A=\pi$ (the polynomial is a symmetric function of $\Delta$). Finally, we find that {\it first-order detuning-modulated CPs of arbitrary lengths $N$} with sing-altenating detunings are implemented for $\Delta$ given by the largest (in absolute value) root of the polynomial Eq. \eqref{poly}. 

As an example, we consider the shortest detuning-modulated composite pulse which has only 2 constituent pulses with durations $\delta t_i = \pi /\sqrt{\Omega^2 + \Delta_i^2}, \; (i=1,2)$. Eq. \eqref{poly} is simply $\Omega^2 - \Delta^2=0$ whose roots $\Delta = \pm \Omega$ realize the shortest broadband sequence. Similarly, given the general solution in Eq. \eqref{poly} it is easy to find sequences of any length $N$ and in the table below we present several examples for $N=3$, 4, and 5. 

\begin{center}
\begin{tabular}{ c c c c } 
 \multicolumn{4}{c}{First-order detuning-modulated CPs} \\
 \hline \hline
 $N$ &&& $\pm (\Delta, -\Delta, \Delta, -\Delta, \dots )/ \Omega $\\ 
 \hline 
 2 && & $ (1, -1) $ \\ 
 3 & &&$ (1, -1, 1) \, \sqrt{3} $ \\ 
 4  & && $(1, -1, 1, -1) \, 2.4142 $ \\
5 && & $(1, -1,1,-1,1) \, 3.0776 $ \\ 
 \hline
\end{tabular}
\end{center}

{\it Second-order composite pulses: Anti-symmetric $\Delta$.} To achieve higher fidelity of the CPs we need to additionally nullify the fourth derivative of the off-diagonal propagator element. We consider odd composite sequences, $N= (2n+1)$, and detuning values that are equal and anti-symmetric with respect to the length of the pulse. That is, $\Delta_i = - \Delta_{N+1-i} \equiv \Delta$, while the detuning of the middle pulse is $\Delta_{n+1}=0$. This anti-symmetric arrangement fulfills the CPT condition Eq. \eqref{cpt3} automatically and similarly to the first-order CPs, the second derivative is zero as it is proportional to the diagonal element of the propagator. For second-order CPs we need to find the detuning values that nullify the fourth derivative and minimize the sixth. This task is considerably simpler as compared to when we do not adopt the anti-symmetric detunings rule. In the table below we present the shortest second-order detuning-modulated CP sequence of length $N=3$ and also the sequences with $N=5$, 7 and 9. We note that these results can easily be extended to large odd lengths $N$.

\begin{center}
\begin{tabular}{ c c c c } 
 \multicolumn{4}{c}{Second-order detuning-modulated CPs} \\
 \hline \hline
 $N$ &&& $\pm (\Delta, -\Delta, \dots, 0, \dots, \Delta, -\Delta) /\Omega $\\ 
 \hline 
 3 & &&$ (1, 0, -1) \, 2.5425 $ \\ 
 5  & && $(1, -1, 0, 1, -1) \, 5.09027 $ \\
7 && & $(1, -1, 1, 0, -1, 1, -1) \, 7.6375$ \\ 
9 &&& $(1, -1, 1, -1, 0, 1, -1, 1, -1) \, 10.1845 $ \\
 \hline
\end{tabular}
\end{center}

The fidelities and infidelities in log scale of the first- and second-order detuning-modulated composite sequences as a function of the individual pulse area error $\delta A/A$ are shown in Fig. \ref{fig2}. For easy reference we also plot the fidelity and infidelity of a resonant pulse and the QI gate error threshold of $10^{-4}$ \cite{QuantumInfo1}. The presented composite sequences are very robust against pulse area errors and the infidelity of the population transfer is well below the QI benchmark even for $\delta A/A$ larger than $10^{-1}$ as compared to less than $10^{-2}$ for single-pulse resonant excitation. We achieve almost an order of magnitude increase in the error tolerance by adding only one additional pulse (first-oder CP), and 1.5 orders of magnitude increase by adding two pulses (second-order CP). Note that the first-order pulses in Fig. \ref{fig2} (green and red curves) have similar robustness as they nullify the second derivative of the propagator element, while the second-order pulse with $N=3$ (blue curve) nullifies the fourth derivative as well. The pulse overhead for our pulses scales as $N$ which is significantly better than that of previous proposals \cite{VitanovSmooth,Torosov:adiabatic,GenovPRL}, where it is $2N$ for $N$ derivatives nullified. This makes our sequences preferable for applications in QIP. In our analysis we also allowed for Gaussian errors of 10\% in the individual pulse lengths and averaged over 100 times (with dashed curves in the bottom frame). We note that the error correction of our CPs is largely unaffected by such inaccuracies. 

\begin{figure}[tbp]
\includegraphics[scale=0.47]{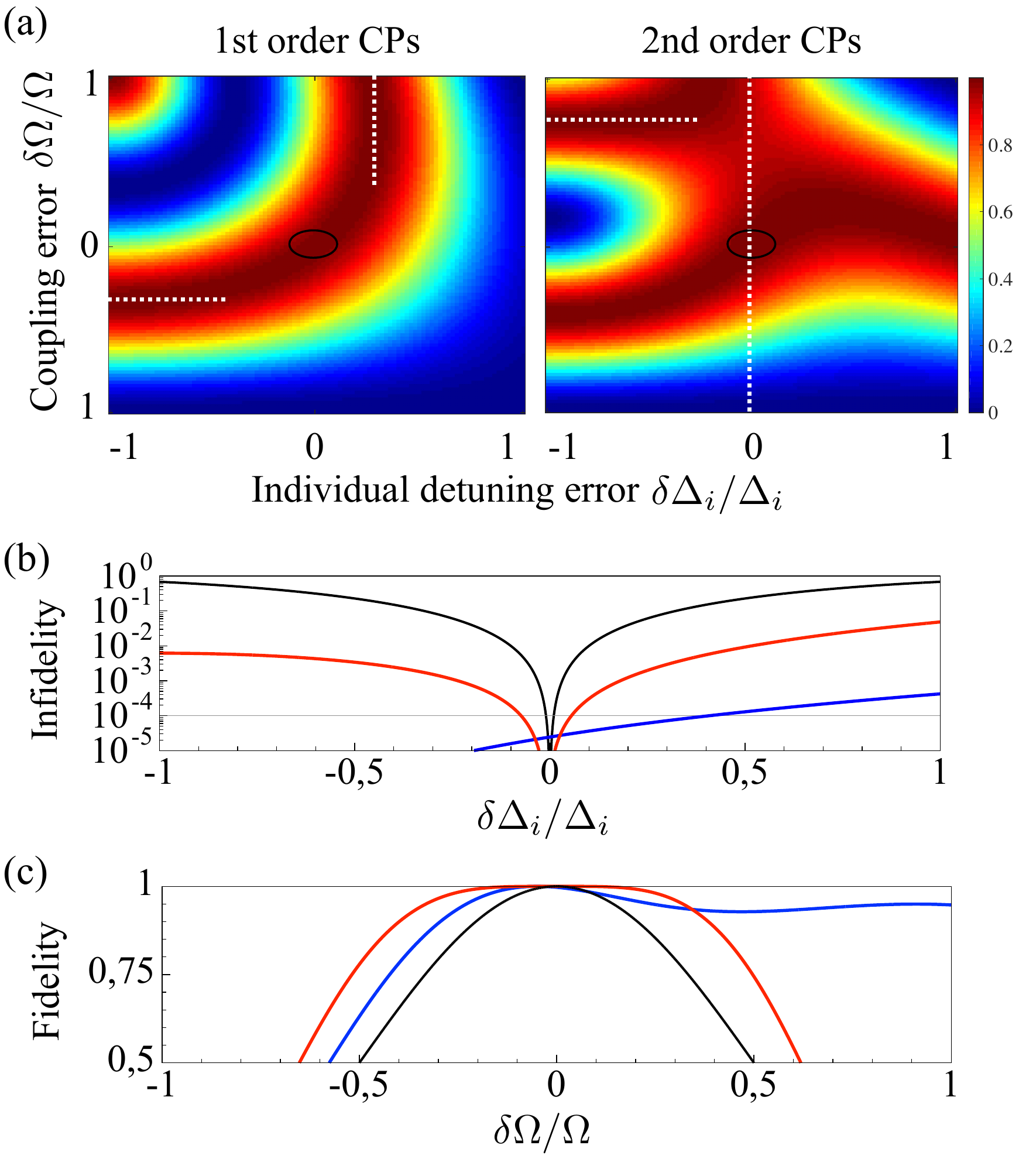}
\caption[justification=justified]{(color online) Robustness of first- and second-order detuning-modulated sequences vs errors in the coupling and detuning. (a) Contour plots of the fidelity of CPT for $N=3$ sequences as a function of errors in the detuning $\delta \Delta_i/\Delta_i$ and the coupling $\delta \Omega/\Omega$. 
(b) Horizontal cut lines from (a) for first-order (red curve) and second-order (blue curve) sequences. (c) Similar to (b) vertical cut lines from (a). For easy reference, the fidelity of a single resonant pulse is shown in black.} 
\label{fig3}
\end{figure}

We examine the robustness of the detuning-modulated composite sequences vs errors in the Rabi frequency and target detuning values. In Fig. \ref{fig3} (a) we present the contour plots of CPT for first-order (left) and second-order (right) CPs of length $N=3$. We note that the contour plots for any other pulses from their respective families look similar. For comparison, we show with black ovals the area with fidelity above $90 \%$ (dark red) of a resonant pulse. By design, our pulses are robust to inaccuracies in the coupling and detuning parameters thus, the area of the parameter space where the fidelity is above 90\% is increased significantly. Furthermore, from the contour plots we find areas in the parameter space (horizontal dashed white lines) where errors in $\delta \Delta_i/\Delta_i$ leave the population fidelity largely unaffected. We plot the horizontal cut lines in log scale in Fig. \ref{fig3} (b) for the first- and second-order CPs (red and blue curves, respectively) compared to the resonant profile shown in black. 
Similarly, in Fig. \ref{fig3} (c) we present the vertical cut lines from (a) that exhibit notable stability against errors in the coupling strength, $\delta \Omega/\Omega$. The presented analysis shows that the derived detuning-modulated sequences are a powerful tool for robust qubits inversion in the presence of any experimental parameter deviation, e.g., pulse duration, pulse amplitude, unwanted detunings, Stark shifts, unwanted frequency chirp, etc.

{\it Realization in coupled waveguides.}
The detuning-modulated CPs, which allow for remarkable error tolerance in qubit inversions, are perfectly suited for implementation in integrated photonic circuits. These pulses offer a unique solution to overcome inaccuracies in fabrication and provide a path for true high-fidelity quantum information processing schemes. In the following, we provide a detailed implementation of our CPs in directional couplers, which serve as a major building block for QIP. In Fig. 1(b), two evanescently-coupled optical waveguides are shown set apart at a constant distance $g$ measured from their centerlines. Within the coupled-mode theory approximation \cite{DirectionalWGs1}, the amplitudes of the fundamental modes in the waveguides obey an equation analogous to Eq. \eqref{Sch} where the coupling is $\Omega = ae^{-bg}$ with $a$ and $b$ being parameters that depend on the material and geometry. As $g$ is fixed, $\Omega$ is also fixed during the entire evolution. The system is on resonance if the waveguides have identical geometry, otherwise there is a real-valued phase mismatch equal to the difference between their respective propagation constants $\beta_{i}$, i.e., $\Delta = (\beta_{1}-\beta_{2})/2$. Thus, it is straightforward to implement our sequences by sequentially changing the waveguides' relative widths such that there are phase jumps in the values of $\Delta$. 

\begin{figure}[tb]
\includegraphics[scale=0.245]{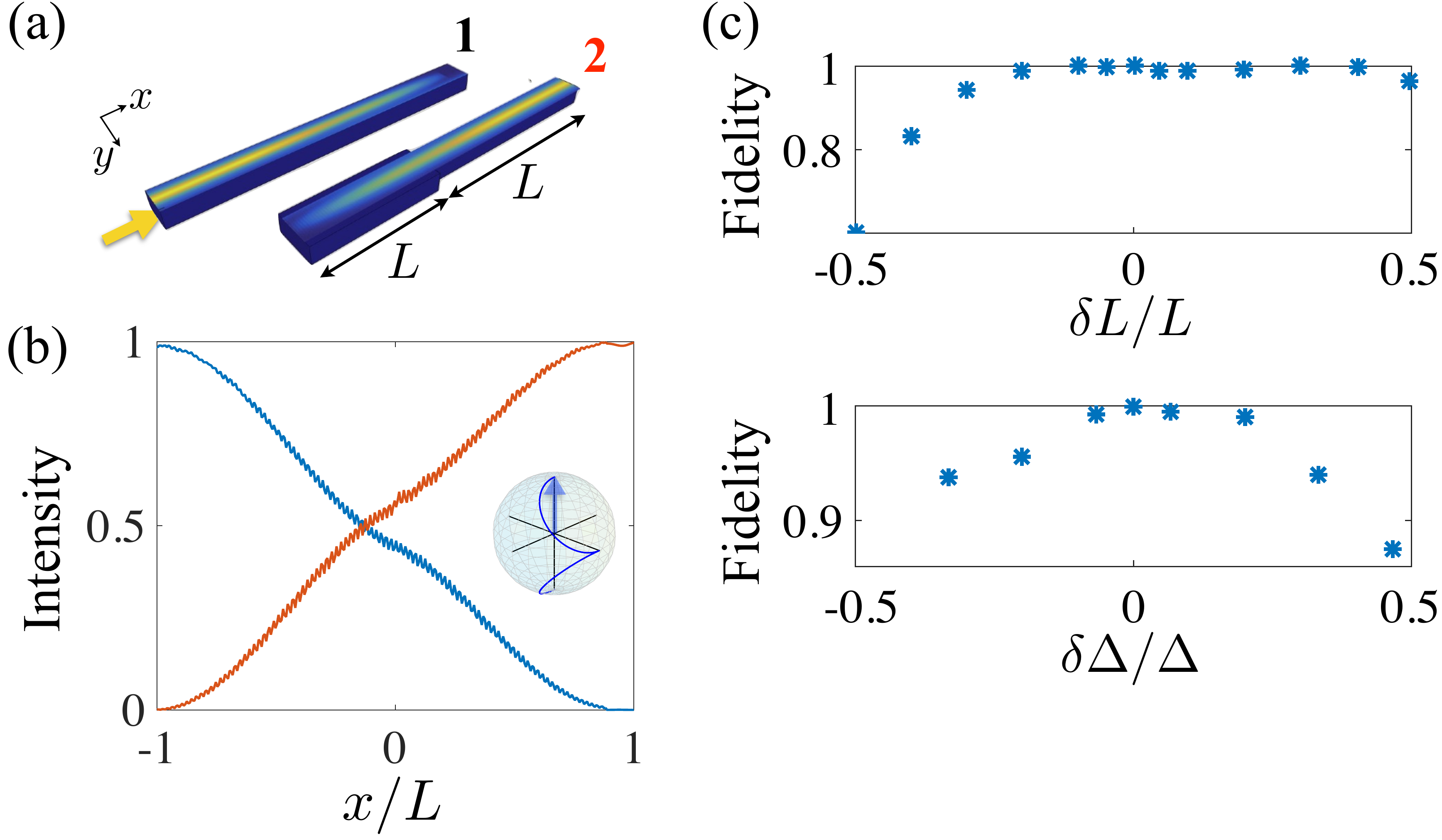}
\caption{Complete light transfer in first-order $N=2$ detuning-modulated composite coupled waveguides. 
(a) An out-of-scale schematic with EME calculation. (b) The light intensity vs the normalized propagation length. (c) Fidelity of the light transfer vs errors in the target phase mismatch and propagation length values.
\label{Figure:Figure4}}
\end{figure}

We present an out-of-scale schematic of the first-order detuning-modulated CP with $N=2$ in coupled waveguides in Fig. \ref{Figure:Figure4} (a). The width of waveguide 1, $w_1$, is fixed, while the width of waveguide 2 changes midlength from $1.034 w_{1}$ to $0.966 w_1$, which realizes the required phase mismatch change in the specific Si on SiO$_2$ configuration that we considered. We perform a calculation by employing an eigenmode expansion (EME) solver to simulate the light propagation along the two waveguides with total length $2L$. We plot the light intensity of waveguides 1 and 2 in Fig. \ref{Figure:Figure4} (b) along with the corresponding Bloch sphere path of the state vector. We observe a complete light switching at the end of the coupler and study its robustness as a function of the phase mismatch error $\delta \Delta /\Delta$ and the propagation length error $\delta L/L$ as compared to their target values. We shown the results in Fig. \ref{Figure:Figure4} (c) and observe high fidelity light transfer even in the presence of errors in excellent agreement with the theoretical calculations (Figs. \ref{fig2} and \ref{fig3}). 

\begin{figure}[tbp]
\includegraphics[scale=0.25]{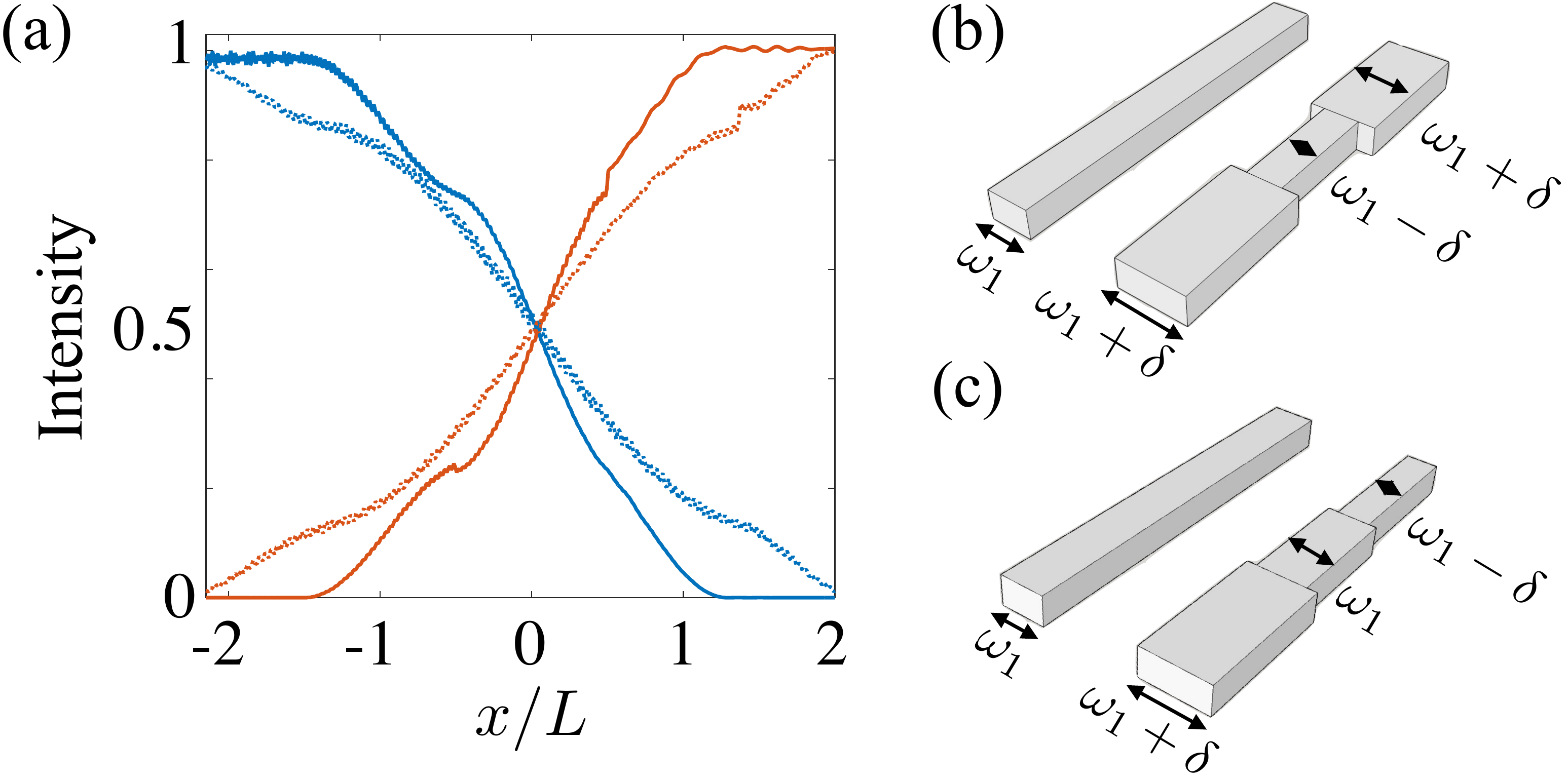} 
\caption{Realizations of first- and second-order detuning-modulated CPs with $N=3$ in coupled waveguides. (a) The light intensity vs the normalized propagation length for a (solid lines) first-order sequence and (dashed lines) second-order sequence. (b) An out-of-scale schematic for the first-order sequence. (c) Same as (b) for the second-order sequence. 
\label{Figure:Figure5}}
\end{figure}

We present the implementation in coupled waveguides of the first- and second-order detuning-modulated CPs with $N=3$ in Fig. \ref{Figure:Figure5}. The length of the separate propagation regions are calculated according to $l=\pi/\sqrt{\Omega^2+\Delta^2}$. For the first-order CP implementation Fig. \ref{Figure:Figure5} (a) these are equal while the width of waveguide 2 changes as $1.057w_1$, $0.943 w_1$, and $1.057w_1$. In Fig. 5(b) we show the second-order CPs implemented where the width of waveguide 2 changes as $1.107 w_1$, $w_1$ and $0.893 w_1$. In a similar fashion, any first-order and second-order composite sequence can be easily realized in a directional waveguide coupler for high-fidelity robust light switching.

{\it Conclusions.} We introduced a novel set of detuning-modulated composite pulse sequences that are characteristically robust to inaccuracies in the systematic parameters including duration, coupling strength and off-resonance errors of the interaction. The control knobs, which we utilized to achieve broadband population inversion, are the detuning parameters of the constituent pulses, while we allow for real coupling constants. We achieved an inversion gate fidelity above the QI threshold vs errors of several percents in the pulse area for a sequence of only 2 constituent pulses, and vs errors of over 10\% for 3 constituent pulses. Furthermore, we showed that the presented sequences are inherently robust to errors in the detuning and coupling parameters. The presented composite pulses are radically different compared to existing composite sequences, which assume complex coupling parameters and modify their phases. Thus, we believe that our general solutions will be the cornerstone for any quantum information protocols and in particular well-suited for practical realization of high-fidelity quantum computing in integrated photonic circuits.  

\begin{acknowledgements}
E. K. acknowledges financial support from the European Union's Horizon 2020 research and innovation programme under the Marie Sk\l odowska-Curie grant agreement No 705256 - COPQE. H. S. is supported by ERC-StG MIRAGE 20-15 project.
\end{acknowledgements}

\bibliography{detuningcp} 

%
%
\end{document}